December 30, 2021

# When bubbles are not spherical: artificial intelligence analysis of ultrasonic cavitation bubbles in solutions of varying concentrations

Ilya Korolev [a, 1], Timur A. Aliev[a], Tetiana Orlova[a], Sviatlana A. Ulasevich[a,*], Michael Nosonovsky[a,*] Ekaterina V. Skorb [a,*]

[a] Infochemistry Scientific Center, ITMO University, 9 Lomonosov St., St. Petersburg, 191002, Russia

**Abstract**

Ultrasonic irradiation of liquids, such as water-alcohol solutions, results in cavitation or the formation of small bubbles. Cavitation bubbles are generated in real solutions without the use of optical traps making our system as close to real conditions as possible. Under the action of the ultrasound, bubbles can grow, oscillate, and eventually, they collapse or decompose. We apply the mathematical method of separation of motions to interpret the acoustic effect on the bubbles. While in most situations, the spherical shape of a bubble is the most energetically profitable as it minimizes the surface energy, when the acoustic frequency is in resonance with the natural frequency of the bubble, shapes with the dihedral symmetry emerge. Some of these resonance shapes turn unstable, so the bubble decomposes. It turns out that bubbles in the solutions of different concentrations (with different surface energies and densities) attain different evolution paths. While it is difficult to obtain a deterministic description of how the solution concentration affects bubble dynamics, it is possible to separate images with different concentrations by applying the Artificial Neural Network (ANN) algorithm. An ANN was trained to detect the concentration of alcohol in a water solution based on the bubble images. This indicates that Artificial Intelligence (AI) methods can complement deterministic analysis in non-equilibrium, near-unstable situations.

[1] Corresponding authors: saulasevich@itmo.ru, nosonovsky@infochemistry.com, skorb@itmo.ru

**Significance statement.** Water-alcohol solutions of various concentrations are studied by irradiating them with ultrasound acoustic waves. The ultrasound irradiation results in the formation of cavitation bubbles, possessing certain natural frequencies of vibration. Near the resonance, the shape of the bubbles is unstable, and it deviates from the spherical shape. Traditional deterministic methods of analysis often fail to study unstable behavior, since small fluctuations grow exponentially and result in large variations of behavior. However, machine-learning methods allow establishing correlations that fail from the traditional methods. Thus, bubbles in solutions of different concentrations behave differently, and the difference is captured by an Artificial Neural Network. Therefore, a novel approach to the analysis of solution concentration has been developed.

## Introduction

Acoustic treatment of liquids is widely used for such purposes as aeration, vibration-induced phase separation, and acoustic spectral analysis of fluid in a flow [1-4]. During ultrasonic acoustic cavitation, small bubbles in liquid are produced by propagating high-frequency pressure fluctuations [5-10]. Newly nucleated bubbles undergo a certain evolution. They often tend to grow, while their size oscillates due to the acoustic excitation causing compressive and tensile stress. During the compressive phase, bubbles shrink, while during the tensile phase they expand for the amount that exceeds shrinking. Consequently, the average bubble radius is growing [10]. Following the oscillation stage, bubbles often either decompose into parts or they can destabilize and completely collapse. The bubble collapse is a very rapid process, which results in high velocities of the liquid-gas interface often leading to shock waves (i.e., exceeding the speed of sound in the fluid) and to very high local temperatures (estimated as high as 5000 K) and high pressures, as well as to the fluorescence [6-10]. This makes ultrasound useful for the sonochemical treatment of solid materials immersed in water including nanostructuring [10-11].

The Rayleigh natural frequency of a bubble, $\omega^2 = \frac{12\gamma}{\rho R_0^3}$, depends on its radius, $R$, liquid density, $\rho$, and the surface energy of the liquid-gas interface, $\gamma$ [12]. Consequently, to a first approximation, the response of a bubble to acoustic vibration is defined by these parameters with the natural frequency decreasing with growing radius. Normally, growing bubbles remain spherical since the spherical shape minimizes the surface area, and, consequently, minimizes the energy of the liquid-gas interface. However, when the frequency of the external acoustic oscillations becomes proportional to the natural frequency of the bubble, resonance can occur, which makes the situation much more complex deviating from the spherical shape. At resonance, the bubble can attain a shape corresponding to a vibrational mode of a droplet, with the growing amplitude of the standing wave [13-14]. The growing amplitude may cause bubble collapse. On the other hand, since the average radius of the bubble continues growing and therefore the natural frequency decreases, the resonance mode shape may disappear. For solutions, such as water-alcohol or salt solutions in water, the evolution may depend on composition, which affects the surface tension, density, and other properties.

Ultrasound has found its application in many areas of science and industry [15], such as the homogenization of food products [16], nanostructuring of surfaces [10-11], and the organic synthesis of compounds [17]. Droplets of different shapes [18-19], as well as colloidal particle clusters [20], have been described in the literature, as well as the acoustic effect on wetting [21-22]. In the present work, we will show the non-trivial application of ultrasound for the qualitative and quantitative determination of the composition of water solutions. For this purpose, we have studied the unstable, close to resonance, the behavior of the cavitation bubble. For the study of the near-resonance unstable behavior, traditional deterministic methods may not work, and statistical methods are needed. The Artificial Intelligence (AI) methods of image recognition, such as Artificial Neural Networks (ANN), are widely used for the classification and clustering of data points and images. These methods can also be used for the analysis of the composition of various solutions and mixtures based on their visual performance (**Fig. 1**).

This article will discuss the evolution of bubbles in water-alcohol solutions and their deviation from the standard spherical shape, including the physical mechanisms of such evolution and their applicability for the visual classification of solutions using artificial intelligence methods. The novelty of our work is that we investigated the cavitation bubble generated in real solutions without the use of optical traps. Thus, our system is as close to real conditions as possible.

## 2. Results and discussion

In most previous studies [12, 23-26], the dynamics and evolution of the cavitation bubble were studied in a ring of light, which serves as an optical trap. In our research, we have used the system as close as possible to real conditions. The motion of the cavitation bubble was not limited, and the bubbles were recorded moved directly in the process of their movement in the media.

### *2.1. Bubble evolution in water-ethanol solutions.*

The bubble cavitation study was performed in a Petri glass dish with a diameter of 10 cm. The video of bubble growth was recorded using a high-speed camera Phantom Miro C110 as shown in **Figure 1.** The Petri dish was placed on an optical Mikmed-6 microscope (LOMO, Russia) equipped with 10× objective, and ultrasound bubbles were generated by the UZG 55-22 ultrasound generator with a resonance frequency 22 ± 3 kHz. The sonotrode was placed at an angle of 45 degrees relative to the surface of the Petri dish. The cavitometer tip was placed constantly in the same position at a distance of 2 cm from the sonotrode.

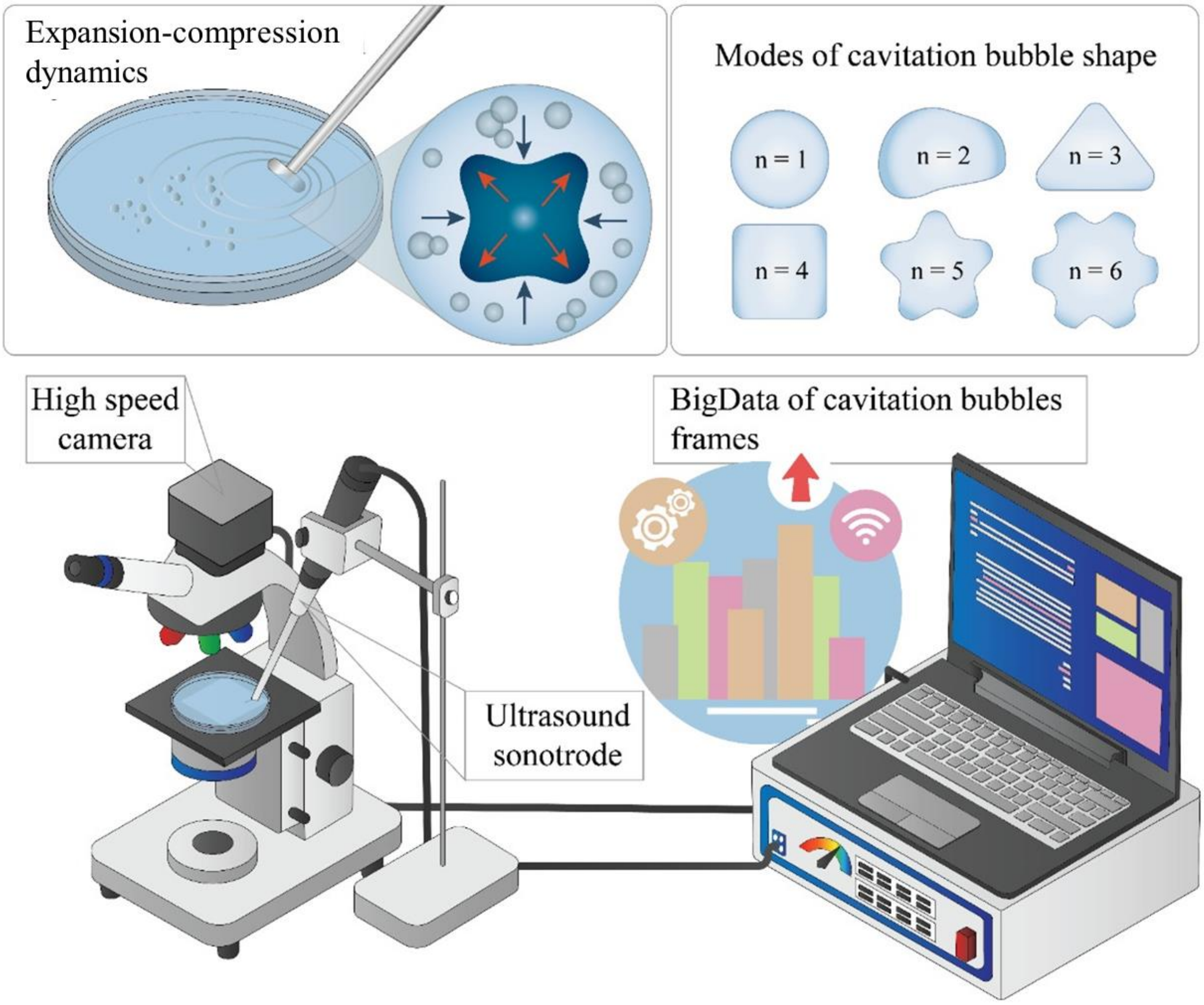

**Figure 1.** The schematic of the experiment set up. The expansion-compression dynamics of the bubbles is captured by a high-speed camera, the frames are analyzed with an Artificial Neural Network.

**Figure 2** shows the change in the dynamics of the cavitation bubble in aqueous-alcoholic solutions containing 5 ÷96 wt. % of ethanol. The shape and radius of bubbles increase with time, which correlates with the existing literature data. The maximum radius of the cavitation bubble appears in a pure water, 25 wt.%, and 50 wt.% ethanol solutions (**Figures 2, 3**).

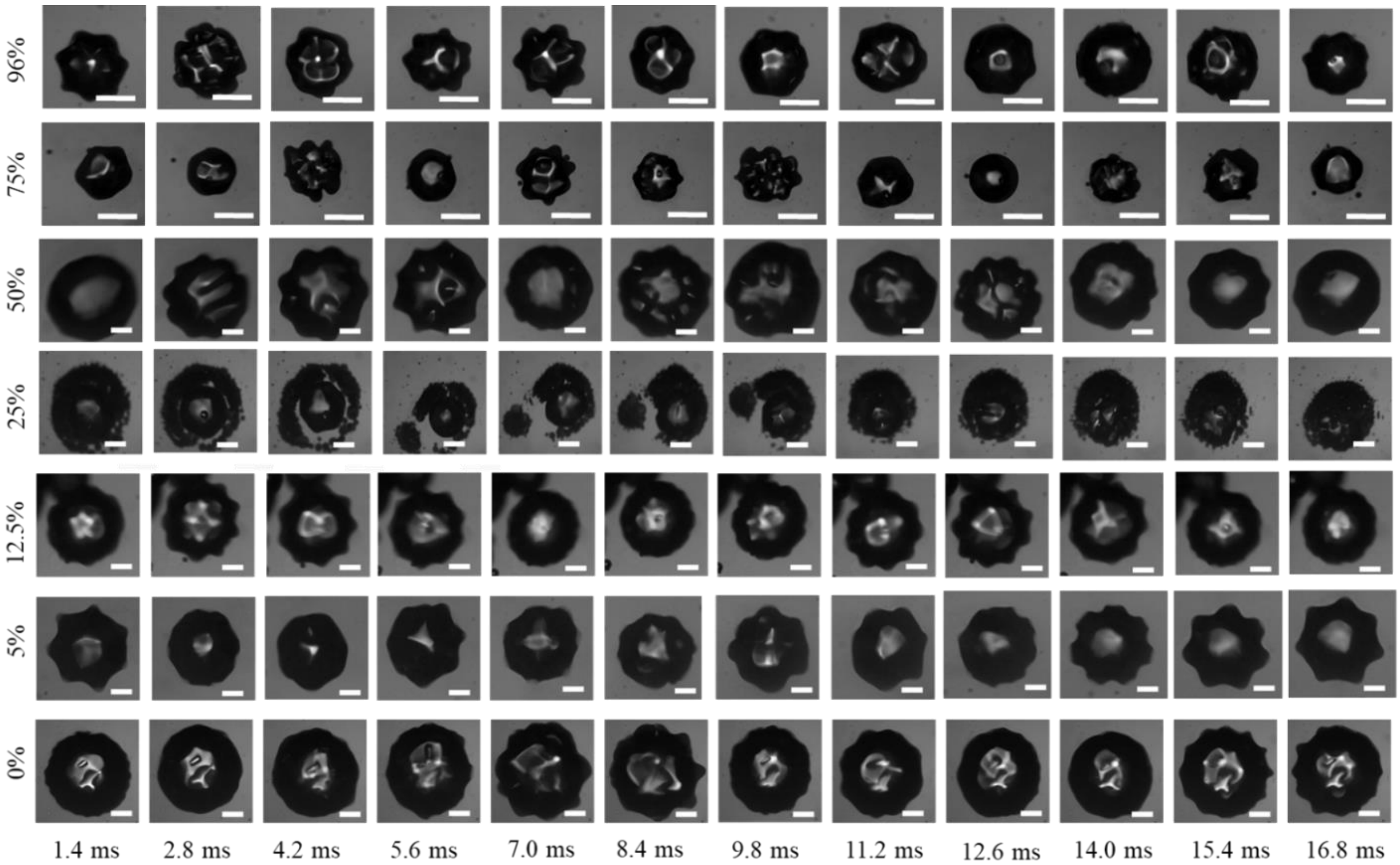

**Figure 2.** Shape oscillations of cavitation bubbles in distilled water (written as 0%) and water-ethanol solutions containing 5 ÷96 wt.% of EtOH. The period of the ultrasonic driving is T = 1/f; the period of the shape oscillation is 31T corresponding to 1.4 ms. Examples of shape oscillations of bubbles are observed in different electrolytes, ethanol mass concentration is marked in Figure as wt.%. The measurements were made using a high-speed camera Phantom Miro C 110. The dynamic of bubble deformation was recorded at 700 frames per second. The images resolution was 768×768 pixels. The image and video analysis were made using Phantom CV 3.3 application. The scale bar is 500 µm.

The behavior of cavitation bubbles and oscillations of their areas were different in solutions with different ethanol concentrations (**Fig. 3**). In particular, the maximum values of the cavitation bubble area are observed in 96%, 75%, and 25% ethanol solutions. The modes of the cavitation bubble increase on increasing the duration of the cavitation.

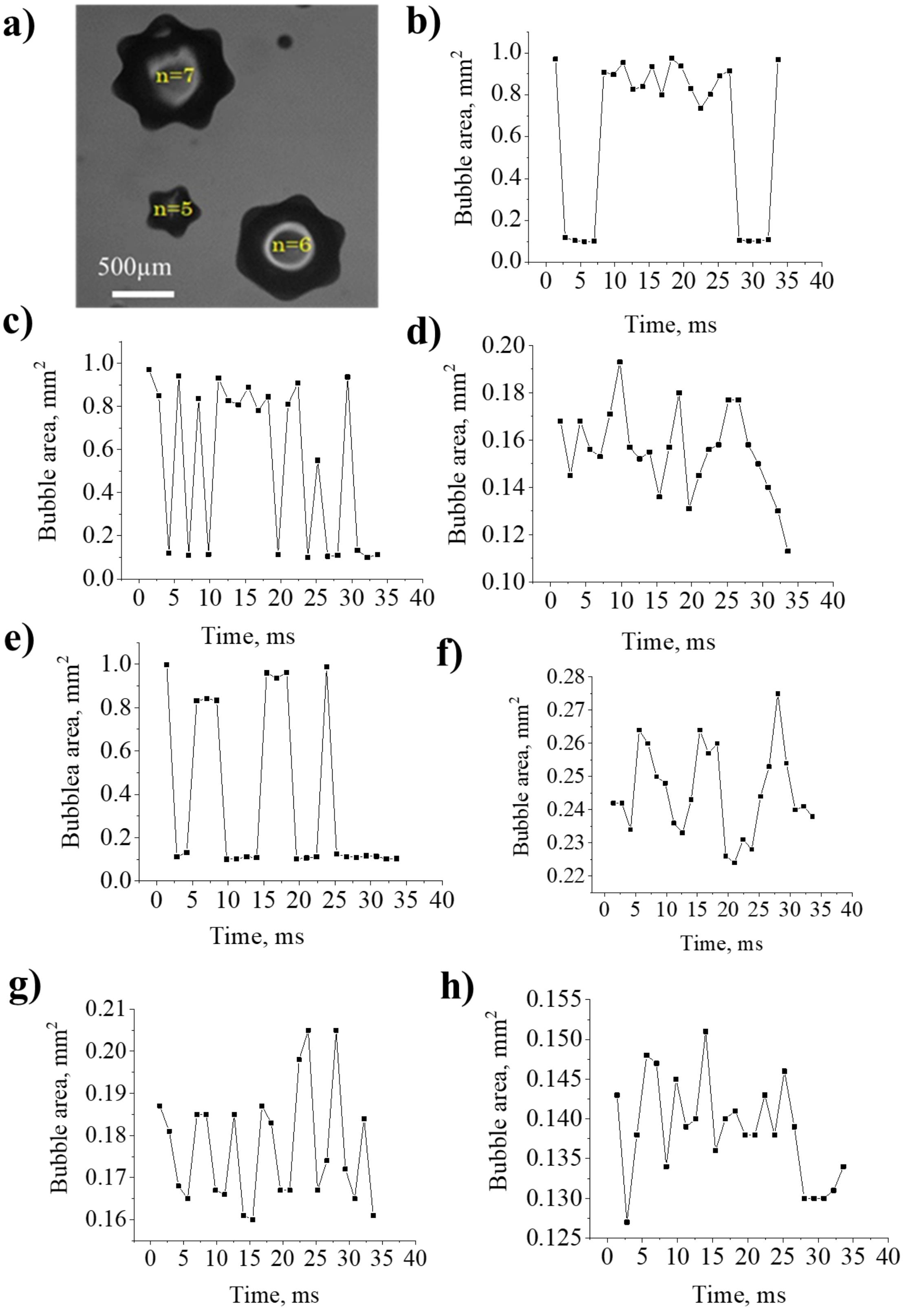

**Figure 3.** a) Image of the cavitation bubble modes; b-h)Temporal oscillations of the cavitation bubble area in water-ethanol solutions at different concentrations: 96 wt.% (b), 75 wt.% (c), 50 wt.% (d), 25 wt.% (e), 12.5 wt.% (f), 5 wt.% (g) and pure water (h). Fast Fourier transform of the cavitation bubble in the water-ethanol solutions.

To measure the effectiveness of cavitation, a cavitometer was used. This device captures vibrations in a medium and converts them into an electrical signal (**Figure 4**). It was observed that the mode of a cavitation bubble can change with varying ultrasonic power. The same tendency was observed for cavitation efficiency. Frequently encountered modes of a cavitation bubble in solutions of alcohol of different concentrations at average values at the same power indices of ultrasound.

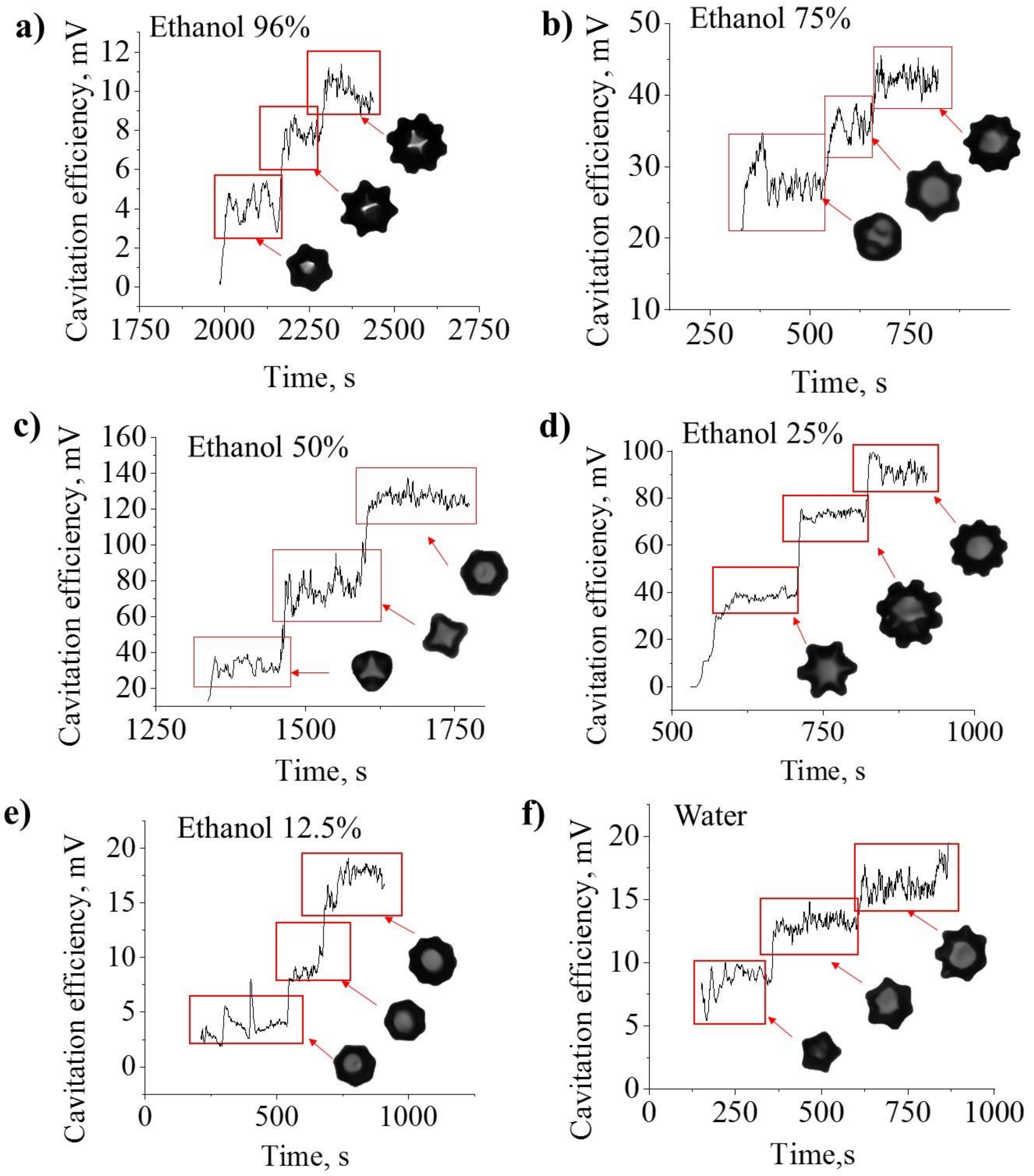

**Figure 4.** (a – f)Temporal evolution of the transient cavitation activity during the sonication of water-ethanol solutions corresponding to the power intensity increase (the first step is 65 W/cm$^{2,}$ the second step is 83 W/cm$^{2}$ , the third step is 100 W/cm$^{2}$ ) in water-ethanol solutions at different concentrations. The inserts show the photo of the most stable shapes of the bubble. The cavitometer was mounted for all experiments in the same position.

Although the cavitometer was positioned at the same place in all experiments, the measured cavitation efficiency showed that the maximum cavitation efficiency was observed for solutions containing 25% and 50% ethanol, respectively.

**Theoretical**

*Bubble nucleation, growth, and collapse.* The dynamic behavior of a spherical bubble with the radius $R$ is governed by the Rayleigh–Plesset equation

$$R\frac{d^2R}{dt^2}+\frac{3}{2}\left(\frac{dR}{dt}\right)^2+\frac{4\nu}{R}\frac{dR}{dt}+\frac{2\gamma}{\rho R}+\frac{\Delta P}{\rho}=0 \quad (1)$$

where $\rho$ is the liquid density, $\nu$ is the kinematic viscosity, $\gamma$ is the surface tension, and $\Delta P$ is the pressure difference inside and outside the bubble. The static equilibrium radius of the bubble is supplied by the Laplace equation, $R_e=\frac{2\gamma}{\Delta P}$, which is obtained from Eq. 1 by setting time derivatives to zero.

Both bubble nucleation and bubble collapse are metastable processes in the sense that the state of the bubble is separated by a small energetic barrier from the liquid state. The Laplace equation defines the energy barrier for the nucleation of new bubbles. For small bubbles, decreasing surface energy (which is proportional to $R^2$) exceeds energetic benefits of the formation of the gas phase (which is proportional to the volume or $R^3$).

$$E=4\pi R^2\gamma+\frac{4}{3}\pi R^3\Delta P \quad (2)$$

The stable state of the minimum energy corresponds to

$$\frac{dE}{dR}=8\pi R\gamma+4\pi R^2\Delta P=0 \quad (3)$$

and the condition is satisfied by the Laplace equation. According to the nucleation theories, the critical radius above which the bubble can spontaneously grow, $R_{cr}=\frac{2\gamma}{\Delta p}$, corresponds to the difference of the saturated vapor pressure and the actual liquid pressure $\Delta p$ [27-28]. Bubbles smaller than a critical radius are energetically unprofitable, because the surface area term prevails over the volumetric term. Larger bubbles can grow when liquid is subjected to tensile stress.

When a harmonically oscillating ultrasound is applied to liquid, the pressure oscillates $\Delta P=\Delta P_0+A\cos\Omega t$ reaching at minima negative values so that the bubble passes through the tensile and compressive stages. Water phase diagram indicates that at low pressures, including negative pressures at room temperatures, the most stable aggregate state of water is vapor. Despite that, pure water can be superheated or stretched, so that it can withstand quite significant values of negative (tensile) pressure without boiling. This is referred to as a metastable superheated state. Ideally, water can withstand tensile stress on the order of $-10^8$ Pa down to the so-called "spinodal limit" of the water phase diagram [27-28]. However, in most practical situations negative pressure results in heterogeneous bubble nucleation due to the presence of various impurities, which serve as gas phase nucleation sites.

The mathematical method of separation of fast and slow motions has been successfully applied to study the effects of fast oscillations in various physical systems where fast small amplitude oscillations are present [1, 9, 29-31]. In this method, high frequency low amplitude vibrations are replaced by an averaged action of a force or of a potential energy field. For bubbles oscillating in the acoustic field created by ultrasound, the effective energy is averaged over the time period of oscillations (for derivation see Supporting materials)

$$\Pi_{eff} = \int_0^{\frac{2\pi}{\Omega}} \left[4\pi R^2(t)\gamma + \frac{4}{3}\pi R^3(t)(\Delta P_0 + A\cos\Omega t)\right] dt \quad (4)$$

where the dependency $R(t)$ is obtained by integrating Eq. 1. The effective averaged acoustic pressure acting on the bubble can be obtained by differentiating the energy

$$p_{eff} = \frac{1}{4\pi R^2}\frac{\partial \Pi_{eff}}{\partial R} \quad (5)$$

Consequently, the average size of the bubble grows slowly subject to the effective pressure given by Eq. 5. At the same time, the bubbles oscillate about the averaged size with high frequency of the ultrasound.

The destabilization and collapse of large bubbles is a dynamic effect caused by the prevalence of the first two terms in Eq. 1. For large bubbles even a small fluctuation of the radius results in an accelerated decrease of the bubble radius governed by the second term in Eq. 1

$$\frac{d^2R}{dt^2} = -\frac{3}{2R}\left(\frac{dR}{dt}\right)^2 \quad (6)$$

Consequently, with decreasing radius the rate of decrease growth causing an accelerated collapse.

*Vibrational modes and non-spherical bubbles.* The analysis above applies only to spherical droplets, since Eq. 1 implies the spherical shape. For resonance cases, the kinetics of non-spherical droplets is not governed anymore by the minimization of the energy functional given by Eq. 2, as destabilization occurs. The amplitude of an initially small perturbation of the spherical shape grows during periodic compression–expansion through a parametric instability, the bubble undergoes shape oscillations.

For a spherical bubble, 3D mode shapes can be quite complex mathematically in the general case. However, for a planar sound wave, oscillations can be viewed as a 2D problem (a 2D projection of a 3D shape). In that case, a small fluctuation from the spherical shape in the polar coordinate system $(r, \theta)$ is given by

$$r_n(\theta) = a_0 + a_n \cos n\theta \quad (7)$$

where $a_n$ are Fourier series coefficients. The natural frequencies of a spherical harmonic distortion of a droplet or bubble of order *n* (for *n*>1) are given by the Rayleigh formula:

$$\omega_n^2 = (n-1)(n+1)(n+2)\frac{\gamma}{\rho R_0^3} \quad (8)$$

where γ and ρ are the surface tension and density of liquid, and $R_0$ is the radius of the bubble [11]. The parameter $a = \frac{\gamma}{\rho R_0^3}$ defines the value of the lowest resonance frequency (*n*=2) as $\omega_2 = \sqrt{\frac{12\gamma}{\rho R_0^3}}$. Note that the case at *n*=1 corresponds to the displacement of the droplet as whole.

Eq. 8 defines corresponding mode shapes. The resonance frequencies correspond to mode shapes, thus *n*=2 corresponds to the elliptical, *n*=3 corresponds to the triangular, *n*=4 to the quadratic mode etc. In the general case, the dihedral symmetry group $D_n$ corresponds to the frequency $\omega_n$.

For a given frequency *f*, the minimum resonance radius is given by

$$R_2 = \sqrt[3]{\frac{12\gamma}{\rho(2\pi f)^2}} \qquad (9)$$

The relationship of the radii corresponding to the natural frequencies is obtained from Eq. 8 as the series

$$R_3 = \left(\frac{40}{12}\right)^{1/3} R_2, \quad R_4 = \left(\frac{90}{12}\right)^{1/3} R_2, \quad R_5 = \left(\frac{168}{12}\right)^{1/3} R_2, \quad R_6 = \left(\frac{280}{12}\right)^{1/3} R_2$$

$$R_7 = \left(\frac{432}{12}\right)^{1/3} R_2,\ R_8 = \left(\frac{630}{12}\right)^{1/3} R_2,\ R_n = \sqrt[3]{\frac{(n-1)(n+1)(n+2)}{12}} R_2 \qquad (10)$$

There are two theories explaining the resonance of an ultrasonically excited droplet. One views resonance as a result of forced vibrations, i. e., the situation when the frequency of the external force (the ultrasound-induced variation of pressure) coincides with one of the natural frequencies of the bubble, $\omega_n$. This results in the amplitude of the vibrations growing unlimitedly eventually leading to the bubble collapse.

Another theory implies the parametric resonance responsible for bubble growth. Parametric resonance is different from regular resonance in that it is not external force that oscillates with the natural frequency of the system, but some parameters of the system change periodically with the frequency related to the natural frequency. The parametric resonance is often governed by the Mathieu equation, $\ddot{x} + \omega_0(1 + \varepsilon \cos \Omega t)x = 0$, where $\varepsilon$ is a small parameter and $\Omega$ is the frequency of change of the parameters. Liu et al. showed that the dynamic equation of encapsulated microbubbles follows the structure of Mathieu's equation, like a gas bubble [15]. In particular, this applies only to encapsulated bubbles.

Small oscillations of the bubble radius near the equilibrium value $R_0$ can be presented as

$$R(t) = R_0 + r(t)$$

Substituting Eq. 11 into Eq. 1 and taking into account $R_o = \frac{2\gamma}{\Delta P_0}$ yields

$$\frac{d^2 r}{dt^2} + \frac{r}{\rho R_0^2}(\Delta P_0 + A\cos\Omega t) = 0 \qquad (11)$$

Thus the resonance frequency is given by $2\Omega = \sqrt{\frac{\Delta P_0}{\rho R_0^2}}$ or $8\pi^2 f^2 = \frac{\Delta P_0}{\rho R_0^2}$ or $R_0 = \frac{1}{2\pi f}\sqrt{\frac{\Delta P_0}{2\rho}}$

For $\Delta P_0 = 10^5$ Pa, $\rho = 10^3\ kg/m^3$, $f = 2 \times 10^4$Hz we find $R_0$ =5.6×10$^{-5}$ or 56 μm.

At the resonance frequency, the droplet may get the shape corresponding to the mode shape, and the amplitude of vibrations may grow for some time, with growth being sustained by viscosity. This may result in the instability leading to the collapse of the bubble or its disintegration into small parts.

Note that while the stability of the bubble can be estimated from various criteria such as Eqs. 4-5, it is usually very difficult to predict how destabilization would occur. Therefore, statistical methods of analysis may be useful to establish insights.

The ultrasonic cavitation in liquids is caused by the acoustic waves of ultrasonic frequency range (typically in the range of 20 kHz and above, corresponding to the wavelengths of 7.5 cm or less). Such ultrasonic waves can be seen as propagating fast low-amplitude oscillations of pressure and density in the liquid medium. The typical values of the pressure amplitude are in the range of $10^5$-$10^7$ Pa correspond to negative (tensile) pressures in the liquid. According to the water phase diagram, boiling temperature decreases with decreasing pressure. Negative pressure corresponds to the gaseous water state, namely, to water vapor. However, the formation of the new vapor phase involves energy barriers. This is because there is surface energy associated with the vapor bubble of a certain size. Consequently, pure (homogeneous) water can withstand quite significant negative pressures on the order of $10^8$ Pa in the stretched or superheated (e.g., heated above the boiling point) state [27]. In practice, however, water is rarely homogeneous since it contains contamination particles, which become nucleation seeds of the new phase. Heterogeneities at the solid-liquid interface, such as asperities or cavities of the surface roughness, also contribute to the heterogeneous nucleation of vapor bubbles.

The density of the ethanol-water mixture changes from 1000 kg $m^{-3}$ for pure water to 798 kg $m^{-3}$ for pure ethanol, while surface tension changes from 0.072 N/m to 0.023 N/m. Therefore, the value of $\frac{\gamma}{\rho}$ changes from $72 \times 10^{-6}$ $m^3/s^2$ to $23.5 \times 10^{-6}$ $m^3/s^2$. For the ultrasound frequency of $f$=20 kHz or ω=125,600 rad $s^{-1}$ we find the upper and lower value of the boundary $R_2 = \left(\frac{12\gamma}{\rho\omega^2}\right)^{\frac{1}{3}} = \left(\frac{12\times72\times10^{-6}}{15.8\times10^9}\right)^{1/3} = 17.6 \times 10^{-6}\ m$ and $\left(\frac{12\times23\times10^{-6}}{15.8\times10^9}\right)^{1/3} = 12.0 \times 10^{-6}\ m$

For n=5, which is frequent in the videos, R~ 90 μm, which is consistent with the observations.

The evolution of an individual bubble subject to ultrasound can pass several stages (i) nucleation (ii) oscillation and growth (iii) destabilization and collapse or (iv) decomposition into parts.

## 4. Artificial Intelligence image clustering

**Software and hardware.** To create the model, the programming language python 3.7, the conda package manager, the pytorch framework, and the jupyter-notebook command shell were used. Additional modules of the python programming language were also used. Numpy for matrices and general math, matplotlib and seaborn for visualization, pandas for working with databases and creating dataframes, opencv-python for image preprocessing. To train the neural network, a computer with an Intel® Core ™ i5-10600 CPU @ 3.30GHz processor, 32 GB of RAM and a GeForce RTX 2060 video card was used.

**Dataset.** The database was created from 12 videos for each alcohol concentration and consisted of a training, validation and test dataset. Figure 5 (c) For the training dataset, 3000 images were taken for each class; for the validation dataset, 300 images were used per class. The test dataset was shot separately and consisted of 500 images for each class.

**Data preprocessing and augmentation.** The video from the highspeed camera was storyboarded and preprocessed using the opencv-python module. The preprocessing involved converting the image from RGB to black and white, which uses only one color layer, instead of 3 for RGB, to reduce the load on the computer and the ability to process more data at a time. With the help of the torchvision package and its methods, a data augmentation was carried out, which included random horizontal flip, random rotation by 60º. After that, the images were converted into tensors and normalized according to ImageNet standards [0.485, 0.456, 0.406], [0.229, 0.224, 0.225]. Similar transformations were used for test images.

**Creating, training and evaluating model.**

The pretrained VGG16 with batch normalization was used to create the model. All layers except the last one were frozen for transfer learning. The last fully connected layer was replaced with 7 output neurons in accordance with the number of classes. Loss function - CrossEntropyLoss, optimizer - Adam, learning rate - 0.001. The number of epochs is 100. The training time was 15 hours and was carried out using a GPU. Figure 5 (a, b) shows the dependence of the number of epochs on the accuracy and loss function for the training and validation dataset. The accuracy during training was 99-100%, however, the validation dataset sometimes lost its accuracy, which can also be seen on the graph of the loss function. Because of this problem, 100 epochs were used, since after 100 epochs the accuracy of the training and validation dataset stabilized. Also, after 100 epochs, retraining began. Figure 5 (d) shows the normalized confusion matrix. It was obtained using an already retrained model on a test dataset. The accuracy of the test dataset was 89.71%.

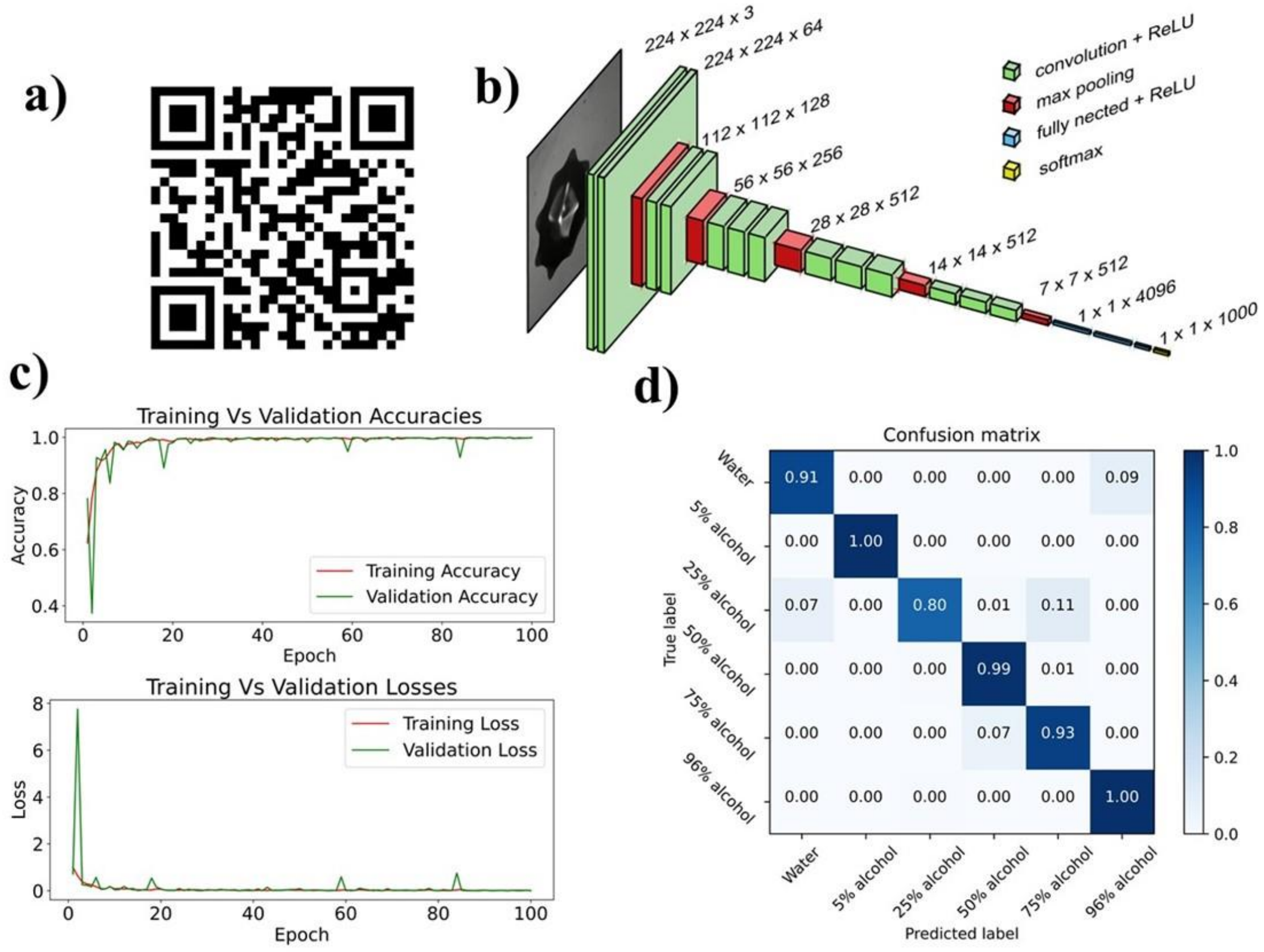

**Figure 5.** Evaluation of the efficiency of a convolutional neural network. **a)** The ratio of accuracy to the number of epochs for the training and validation dataset. **b)** The ratio of the loss function to the number of epochs for the training and validation dataset. **c)** The composition of the dataset, divided into training, validation and testing. **d)** Normalized confusion matrix for the test dataset, the accuracy was 89.71%.

## Conclusion

Ultrasonic irradiation of liquids, such as water solutions, results in the formation of small cavitational bubbles, which grow, oscillate, and eventually collapse. In most situations, the spherical shape of a bubble is the most energetically profitable as it minimizes the surface energy. The effect of the acoustic oscillations can then be presented as an average force leading to the bubble expansion. This force is calculated by the method of separation of motions. However, when the frequency of the acoustic oscillations is in resonance with the natural frequency of the bubble, the Rayleigh–Plesset equation governing the bubble dynamics, is not applicable. In the resonance case, droplets take the shape corresponding to the frequency mode. These shapes are characterized by the dihedral symmetry group, and they can be easily detected visually.

While resonance and stability criteria provide conditions for the destabilization of the bubble, they do not provide insights on the droplet dynamics or kinetics near the unstable states. Such kinetics should be viewed as near-critical or as chaotic, with small random fluctuations defining the bubble evolution. To understand this chaotic behavior, statistical methods can be used. Among such methods is clustering visual images using the Artificial Intelligence algorithms. While it is not possible to obtain a deterministic description of how the concentration of a solution affects bubble dynamics, it is still possible to separate images with different concentrations by applying the Artificial Neural Network algorithm. This demonstrates that AI methods can complement deterministic analysis in the non-equilibrium, near-unstable situations. The model can be applied. For example, to computer vision software, which can be installed on a computer with a microscope to automatically analyze bubbles in real time.

**Acknowledgments**

This work was supported by the Ministry of Science and Higher Education of Russian Federation, goszadanie no. FSER-2021-0013. The infrastructural supported by the Government of the Russian Federation through the ITMO Fellowship and Professorship program.

**References**

1. MS Hasan, M Nosonovsky Method of separation of vibrational motions for applications involving wetting, superhydrophobicity, and microparticle extraction Physical Review Fluids 5 (5), 054201

2. A. Priev and A. Sarvazyan, Cylindrical standing wave resonator for liquid food quality control, J. Acoust. Soc. Am. 125, 2593 (2009).

3. V. Ostasevicius, V. Jurenas, I. Golinka, R. Gaidys, and A. Aleksa, Separation of microparticles from suspension utilizing ultrasonic standing waves in a piezoelectric cylinder actuator, Actuators 7, 14 (2018).

4. C. Devendran, I. Gralinski, and A. Neild, Separation of particles using acoustic streaming and radiation forces in an open microfluidic channel, Microfluid. Nanofluid. 17, 879 (2014).

5. E V. Skorb and D V. Andreeva Surface Nanoarchitecture for Bio-Applications: Self-Regulating Intelligent Interfaces Adv. Funct. Mater. 2013, DOI: 10.1002/adfm.201203884

6. E V. Skorb, D V. Andreeva, and H Mçhwald, Generation of a Porous Luminescent Structure Through Ultrasonically Induced Pathways of Silicon Modification Angew. Chem. 2012, 124, 1 – 6

7. J Gensel, T. Borke, N P Pérez, A Fery, D V. Andreeva, E Betthausen, A H. E. Müller, H Möhwald, and E V. Skorb Cavitation Engineered 3D Sponge Networks and Their Application in Active Surface Construction Adv. Mater. 2012, 24, 985–989

8. E V. Skorb, Helmuth Möhwald Ultrasonic approach for surface nanostructuring Ultrasonics Sonochemistry 29 (2016) 589–603

9. M Sabbouh, A Nikitina, E Rogacheva, L Kraeva, SA Ulasevich, EV Skorb, M. Nosonovsky 2021, Separation of motions and vibrational separation of fractions for biocide brass, Ultrasonics Sonochemistry 80, 105817

10. E. V. Skorb and D. V. Andreeva Bio-inspired ultrasound assisted construction of synthetic sponges J. Mater. Chem. A, 2013,1, 7547–7557

11. J. Dulle, S. Nemeth, E. V. Skorb, D. V. Andreeva, Sononanostructuring of zinc-based materials, RSC Adv. 2 (2012) 12460–12465. https://doi.org/10.1039/c2ra22200k

12. R. W. S. Rayleigh Proc. R. Soc. London, 1879, 29 , 71 —97.

13. V. Poulichet, Axel Huerre and Valeria Garbin Shape oscillations of particle-coated bubbles and directional particle expulsion, Soft Matter, 2017, 13, 125—13

14. Y. Liu and Q. Wang, Stability and natural frequency of nonspherical mode of an encapsulated microbubble in a viscous liquid, Physics of Fluids 28, 062102 (2016);

15. Honarvar, F., & Varvani-Farahani, A. (2020). A review of ultrasonic testing applications in additive manufacturing: Defect evaluation, material characterization, and process control. Ultrasonics, 106227.

16. Aboutorab, M., Ahari, H., Allahyaribeik, S., Yousefi, S., & Motalebi, A. (2021). Nano-emulsion of saffron essential oil by spontaneous emulsification and ultrasonic homogenization extend the shelf life of shrimp (Crocus sativus L.). Journal of Food Processing and Preservation, 45(2), e15224

17. Shahinshavali, S., Hossain, K. A., Kumar, A. V. D. N., Reddy, A. G., Kolli, D., Nakhi, A., & Pal, M. (2020). Ultrasound assisted synthesis of 3-alkynyl substituted 2-chloroquinoxaline derivatives: Their in silico assessment as potential ligands for N-protein of SARS-CoV-2. Tetrahedron letters, 61(40), 152336

18. S. Guttman, Z. Sapir, M. Schultz, A. V. Butenko, B. M. Ocko, M. Deutsch, E. Sloutskin How faceted liquid droplets grow tails, Proc. Nat. Acad. Sci. 2016, 113 (3) 493-496;

19. R. Tadmor, A. Baksi, S. Gulec, S. Jadhav, H. E. N'guessan, K. Sen, V. Somasi, M. Tadmor, P. Wasnik, and S. Yadav, Drops that change their mind: Spontaneous reversal from spreading to retraction, Langmuir 35, 15734 (2019).

20. BA Grzybowski, K Fitzner, J Paczesny S Granick, From dynamic self-assembly to networked chemical systems, Chem. Soc. Rev. 2017, 46, 5647–5678.

21. E. Y. Bormashenko, Wetting of Real Surfaces (Walter De Gruyter, 2013)

22. O. Manor, Diminution of contact angle hysteresis under the influence of an oscillating force, Langmuir 30, 6841 (2014).

23. Pouliopoulos, A. N., Li, C., Tinguely, M., Garbin, V., Tang, M. X., & Choi, J. J. (2016). Rapid short-pulse sequences enhance the spatiotemporal uniformity of acoustically driven microbubble activity during flow conditions. The Journal of the Acoustical Society of America, 140(4), 2469-2480.

24. Dollet, B., van der Meer, S. M., Garbin, V., de Jong, N., Lohse, D., & Versluis, M. (2008). Nonspherical oscillations of ultrasound contrast agent microbubbles. Ultrasound in medicine & biology, 34(9), 1465-1473.,

25. Dollet, B., Marmottant, P., & Garbin, V. (2019). Bubble dynamics in soft and biological matter. Annual Review of Fluid Mechanics, 51, 331-355.,

26. Baresch, D., & Garbin, V. (2020). Acoustic trapping of microbubbles in complex environments and controlled payload release. Proceedings of the National Academy of Sciences, 117(27), 15490-15496

27. SH Yang, M Nosonovsky, H Zhang, KH Chung Nanoscale water capillary bridges under deeply negative pressure Chemical Physics Letters 451 (1-3), 88-92

28. E. Herbert, F. Caupin The limit of metastability of water under tension: theories and experiments J. Phys.: Condens. Mat., 17 (2005), p. S3597

29. P. Kapitza, Pendulum with a vibrating suspension, Usp. Fiz. Nauk 44,7 (1951).

30. R. Ramachandran, N. Maani, V. L. Rayz, and M. Nosonovsky, Vibrations and spatial patterns inbiomimetic surfaces: Using the shark-skin effect to control blood clotting, Philos. Trans. R. Soc., A374,20160133(2016).

31. R. Ramachandran and M. Nosonovsky, Vibro-levitation and inverted pendulum: parametric resonance in vibrating droplets and soft materials, Soft Matter10,4633(2014).

**Table of content entry**

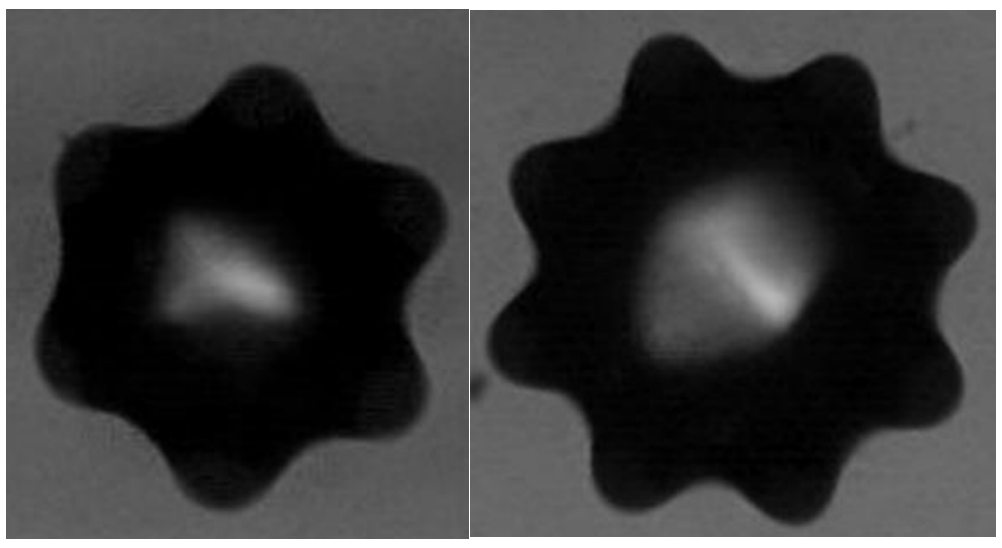

Cavitation bubbles with six and eight edges